\documentclass[conference]{IEEEtran}

\usepackage{bm}
\usepackage{multirow}
\usepackage{float}
\usepackage[cmex10]{amsmath}

\ifCLASSINFOpdf
  \usepackage[pdftex]{graphicx}
\else
\fi
\usepackage{tikz}

\newcommand\copyrighttext{%
  \footnotesize \textcopyright 2026 IEEE. Personal use of this material is permitted. Permission from IEEE must be obtained for all other uses, in any current or future media, including reprinting/republishing this material for advertising or promotional purposes, creating new collective works, for resale or redistribution to servers or lists, or reuse of any copyrighted component of this work in other works.}
\newcommand\copyrightnotice{%
\begin{tikzpicture}[remember picture,overlay]
\node[anchor=south,yshift=10pt] at (current page.south) 
  {\fbox{\parbox{\dimexpr\textwidth-\fboxsep-\fboxrule\relax}{\copyrighttext}}};
\end{tikzpicture}%
}

\begin{document}
\bstctlcite{IEEEexample:BSTcontrol}
\title{RIS-Aided Sensing: Experimental Validation of Radar 3D Imaging in the mmWave Band}

\author{\IEEEauthorblockN{Sergio Micó-Rosa, Alvaro Villaescusa-Tebar, Saúl Fenollosa, Carlos Villena-Jiménez, \\ Monika Drozdowska, and Narcis Cardona}
\IEEEauthorblockA{iTEAM Research Institute\\
Universitat Politècnica de València\\
Valencia, Spain 46022\\
Email: \{sermiro, alvilte1, sjfenarg, cviljim, mdrozdo, ncardona\}@iteam.upv.es}}


\maketitle
\copyrightnotice
\begin{abstract}
%\boldmath
The transition toward 6G networks demands energy-efficient hardware capable of active interaction with the environment. Reconfigurable Intelligent Surfaces (RIS) have emerged as a key technology for Integrated Sensing and Communications (ISAC), enabling geometric environment recognition with minimal power consumption. However, achieving targeted 3D spatial mapping in a fully autonomous, closed-loop system remains a significant challenge. In this work, we validate experimentally an autonomous mmWave 3D imaging framework that integrates an Frequency-Modulated Continuous Wave (FMCW) radar with a 1-bit RIS and a Vector Network Analyzer (VNA) to perform targeted 3D reconstruction. The FMCW radar acts as a coarse localizer, providing real-time spatial priors to define dynamic Regions of Interest (ROI). These coordinates are translated into optimized RIS phase profiles to perform Stepped-Frequency Continuous-Wave (SFCW) measurements. We experimentally validate the system through three diverse scenarios, including metallic mannequins, calibration spheres, and a complex multi-target environment containing human subjects and an Automated Guided Vehicle (AGV). The results demonstrate accurate 3D voxel-based reconstruction of targets even at reduced angular resolutions, advancing the feasibility of RIS-based sensing for industrial and security applications.
\end{abstract}


% \fi
\IEEEpeerreviewmaketitle
\section{Introduction}
The transition towards 6G networks drives the search for energy-efficient hardware capable of redefining how signals interact with the environment. Reconfigurable Intelligent Surfaces (RIS) have emerged as a groundbreaking technology that transforms the propagation medium from a passive channel into an active system component \cite{Basar2019WirelessSurfaces}. Physically, a passive RIS is a planar array of low-cost electromagnetic unit cells that can independently control the phase of incident waves. By electronically tuning these elements, the RIS manipulates electromagnetic wavefronts to perform passive beamforming and localized energy focusing \cite{DiRenzo2020SmartAhead}. While initially intended to enhance communication performance and coverage, the RIS is now recognized as a vital enabler for Integrated Sensing and Communications (ISAC). This architecture offers a versatile platform to embed radar-like capabilities into wireless networks, while addressing the hardware complexity and cost constraints of traditional systems \cite{Magbool2025AOpportunities}. Its ability to shape the channel not only enables data transmission, but also allows geometric information to be extracted from the environment, enabling advanced applications ranging from passive detection of human activity to volumetric imaging and environment recognition with minimal energy consumption compared to traditional radar architectures \cite{Li2025PassiveChallenges}.

Leveraging these beamforming capabilities, recent literature has validated diverse practical use cases for RIS-assisted sensing. In the radar domain, the RIS mitigates physical limitations of conventional hardware, enabling 3D target localization with 2D sensors via virtual link geometry \cite{Liu2024ReconfigurableRadar}, and enhancing angular resolution to distinguish adjacent targets that would be indistinguishable to radar alone \cite{Vejling2025RIS-AssistedSensing}. Beyond static detection, dynamic ISAC architectures now support continuous tracking of multiple users in mobility scenarios \cite{Zhang2022JointInformation}, while self-localization schemes, that have been experimentally validated, allow radar-equipped users to estimate their position and speed using the RIS as a passive reference \cite{Kim2025RIS-EnabledRadar}. Despite this theoretical foundation and preliminary experimental evidence for point-target detection, generating accurate 3D environmental imaging remains a distinct challenge that requires further investigation.

To address the specific demands of high-resolution imaging, the literature has evolved towards sophisticated architectures supported by advanced hardware designs, such as active RIS architectures to compensate for path loss \cite{Sun2024ActiveImaging}, and very-large-scale RIS for dynamic wavefront control \cite{Malevich2024Very-Large-ScaleWaves}. Theoretical frameworks utilize these surfaces to synthesize large virtual apertures, employing techniques like the Virtual Source Principle \cite{Ilgac2025Ris-AidedPrinciple}, Computational Radar Coincidence Imaging \cite{Zirak2025ASurfaces}, or Deep Reinforcement Learning schemes \cite{Hu2021MetaSensing:Learning} to reconstruct high-fidelity environmental maps. From an experimental perspective, significant measurement-based demonstrations have been reported. In particular, mmWave ISAC system demonstrations have validated the feasibility of user detection in real-world scenarios \cite{Li2024UserDemonstration}, while pioneering studies in WiFi imaging have successfully exploited RIS to generate high-resolution two-dimensional images using opportunistic signals \cite{He2023High-ResolutionSurfaces}. Nevertheless, existing solutions typically rely on non-controllable sources, are limited to 2D projections, or focus on classification tasks rather than volumetric spatial mapping. Consequently, the realization of a fully autonomous RIS-aided radar imaging system, which integrates sensing and control in a closed loop to perform targeted 3D reconstruction, remains an open challenge.

In this work, we introduce a novel autonomous imaging framework that establishes a closed control loop between wide-area detection and localized 3D imaging. Departing from existing open-loop strategies that rely on exhaustive codebook scanning, our approach employs a commercial Frequency-Modulated Continuous Wave (FMCW) radar as a primary guide to dynamically control the RIS-aided imaging process. The specific contributions of this work are three-fold: 1) the implementation of a versatile hardware-in-the-loop experimental testbed integrating a 1-bit RIS, a Vector Network Analyzer (VNA) for accurate scene reconstruction acting as an Stepped-Frequency Continuous-Wave (SFCW) radar, and a commercial FMCW radar for real-time target tracking; 2) the development of a spatial mapping algorithm that translates radar-derived coordinates into optimized RIS phase profiles for targeted scanning; and 3) a comprehensive experimental validation demonstrating consistent 3D voxel-based reconstruction. By delegating the search phase to a dedicated radar, we demonstrate that effective imaging is achievable, advancing the feasibility of RIS-based sensing for real-time industrial and security applications.

The rest of this paper is organized as follows. Section~II details the experimental methodology, including the hardware specifications, measurement setup, and defined scenarios. Section~III formulates the system model, covering RIS-aided signal propagation, radar signal processing, and the 3D voxelization framework. Section~IV presents the experimental results, while Section~V provides a critical discussion of the findings. Finally, Section~VI concludes the work.
\section{Methodology}
% just do the begining of methodology
This section describes the methodology used for the measurements. First, the hardware specifications are presented, followed by a description of the measurement setup and parameters. Next, the measurement procedure is explained. Finally, the three measurement scenarios considered in this study are described.
\subsection{Radar description}
The primary detection sensor is a highly integrated commercial FMCW radar transceiver operating in the 60-64 GHz mmWave band (Texas Instruments AWR6843). This sensor serves as the coarse localizer, providing the initial spatial priors required to configure the RIS phase profiles. For this study, the radar is configured to provide distance and azimuth information of detected targets in a coverage of approximately $120^\circ$ in azimuth and $30^\circ$ in elevation. It achieves spatial resolutions superior to 4 cm, with a beamforming-assisted angular resolution of less than $10^\circ$. %Key hardware specifications are summarized in Table~\ref{tab:RadarParams}.
% \begin{table}[h!]
% \centering
% \caption{FMCW Radar Transceiver Key Specifications.}
% \label{tab:RadarParams}
% \renewcommand{\arraystretch}{1.2}
% \begin{tabular}{|l|c|}
% \hline
% \textbf{Parameter} & \textbf{Value} \\ \hline
% \textbf{Frequency Coverage} & 60--64 GHz \\ \hline
% \textbf{Bandwidth} & 4 GHz (Continuous) \\ \hline
% \textbf{Tx Output Power} & 12 dBm \\ \hline
% \textbf{Rx Noise Figure} & 12 dB \\ \hline
% \textbf{Phase Noise} & $-93$ dBc/Hz @ 1 MHz \\ \hline
% \textbf{Architecture} & 3 Tx / 4 Rx Channels \\ \hline
% \textbf{Processing} & C674x DSP + Arm Cortex-R4F \\ \hline
% \end{tabular}
% \end{table}
\subsection{RIS description}
The RIS panel used in this study is FR2 passive dual-polarized RIS fabricated by Greenerwave, whose design process is explained in \cite{Gros2023Design5G/6G}.

The FR2 RIS presents dual-polarization, with two possible states per element and polarization: $0^\circ$ and $180^\circ$ phase shifts. It operates in the 25.5-31.5 GHz frequency band; however, a phase shift of approximately $180^\circ$ between the two element states is only ensured within the 27.5–29.5 GHz frequency range. The overall dimensions of the FR2 RIS housing are $33 \times 27 \times 4$~cm. The housing contains four identical metasurfaces, each measuring $10 \times 10$~cm and arranged symmetrically. Each metasurface consists of 20 × 20 unit cells arranged in a periodic pattern separated by a half wavelength. Thus, the RIS array is $20 \times 20$~cm, and $40 \times 40$ unit cells. In total, the RIS comprises 1600 unit cells and 3200 diodes, enabling dual-polarized operation. The RIS is controlled by an FPGA-based control board connected to a laptop via USB, with control implemented using Python. The RIS pattern is modified based on the algorithm explained in the following sections. The beam steering update rate is more than 1~kHz, reaching up to 75~kHz.
% \begin{table}[h!]
% \centering
% \caption{RIS FR2 panel parameters.}
% \label{tab:RISparams}
% \renewcommand{\arraystretch}{1.2}
% \begin{tabular}{|l|c|}
% \hline
% \textbf{Parameter} & \textbf{Value} \\ \hline
% \textbf{Frequency} & 25.5--31.5 GHz \\ \hline
% \textbf{Dimensions} & 33x27x4 cm \\ \hline
% \textbf{Cells} & 40x40 \\  \hline
% \textbf{Polarization} & dual polarization \\ \hline
% \textbf{Update rate} & >1 kHz, Python drivers \\  \hline
% \end{tabular}
% \end{table}
\subsection{Measurement setup}
Measurements were conducted in a controlled semi-anechoic chamber environment. The setup consisted of the Rohde \& Schwarz ZNB40 VNA, two mmWave horn antennas with a half-power beamwidth (HPBW) of $30^\circ$ and a gain of 16 dBi, Greener Wave FR2 RIS, a FMCW radar, and a laptop that acts as the master controller.

The center of the RIS and both antennas were mounted at a height of 1.3~m. Antennas are vertically polarized, placed close together to emulate quasi-monostatic conditions. The distance between antennas' feeders and the RIS surface is 0.6~m, and they are placed directly in front of the RIS center, along its central axis. Antennas were connected to each of the two ports of the VNA with a 5~m RF cable and two 2~m RF cables, respectively. The setup parameters are summarized in Table~\ref{tab:VNAparams}.
\begin{table}[!t]
\centering
\caption{VNA parameters.}
\label{tab:VNAparams}
\renewcommand{\arraystretch}{1.2}
\begin{tabular}{|l|c|}
\hline
\textbf{Parameter} & \textbf{Value} \\ \hline
\textbf{Frequency} & 26.5--30.5 GHz \\ \hline
\textbf{Transmitting power} & 0 dBm \\ \hline
\textbf{Sweep points} & 256 \\  \hline
\textbf{IF BW} & 1 kHz \\ \hline
\end{tabular}
\end{table}

\begin{figure}[!t]
\centering
\includegraphics[width=0.9\columnwidth]{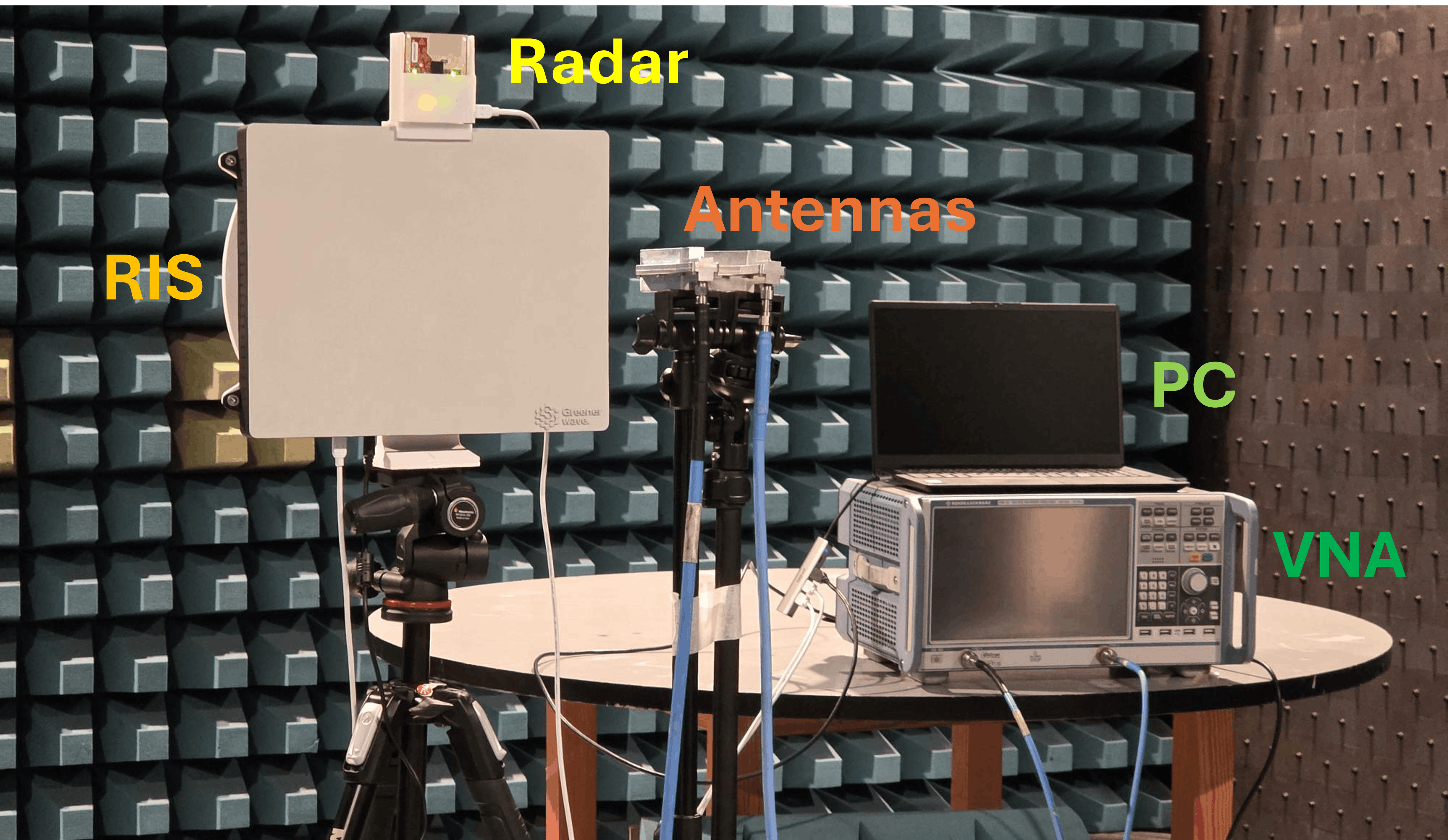}
\caption{Photo of the measurement setup.}
\label{fig:meas}
\end{figure}

The VNA frequency band was selected to cover the bandwidth where the RIS accomplishes the desired $180^\circ$ phase shift between its element states. Hence, the central frequency of 28.5~GHz was selected and 4~GHz bandwidth was used to cover the RIS working frequency and achieve a good spatial resolution of 3.75~cm in the monostatic radar case. The number of points $K$ is selected to be small but sufficient to cover the entire range of the scene, ensuring that no target is missed. This selection takes into account a maximum observable distance of 9.60~m ($256 \times 3.75$~cm). The setup is shown in Fig.~\ref{fig:meas}.

\subsection{Measurements procedure}
% explicar el funcionamiento del algoritmo más detallado, como todo se conecta y funciona
The proposed imaging system operates under an autonomous closed-loop sensing paradigm. The procedure is divided into a dual-step detection and imaging framework orchestrated by a Python-based master control engine.
\subsubsection{Initial Target Detection and ROI Definition}
The sensing cycle begins with the FMCW radar, mounted coaxially with the RIS, acting as the "eyes" of the system. This radar performs wide-area monitoring to estimate the target's spatial coordinates, specifically azimuth ($\phi$) and distance ($R$), in real-time.

Once a target is detected, the control engine dynamically defines a 3D Region of Interest (ROI) centered at the radar-estimated coordinates. The ROI is characterized by an angular span in both $\phi$ and elevation ($\theta$), with a predefined angular step $\Delta \psi$ (different depending on the scenario). This constrained search space prevents exhaustive scanning of the entire hemisphere in front of the RIS, focusing the electromagnetic energy exclusively on the spatial volume occupied by the detected targets.
\subsubsection{RIS Configuration and SFCW Scanning}
For each discrete angular pair $(\phi_i, \theta_j)$ within the ROI, the system executes a measurement loop. The phase-shift configuration for the $N \times M$ elements of the RIS is calculated using the phase-gradient and focusing principles described in \cite{Tang2022PathBand}, detailed in Section~\ref{sec:ris_phaseshift}. The target's position is thus determined using the angular pair from the ROI and the distance provided by the FMCW radar. Therefore, the RIS control algorithm computes the optimal 1-bit phase states ($0, \pi$) required to perform passive beamforming towards the specific 3D point.
\subsubsection{Data Acquisition and VNA Integration}
After updating the RIS hardware with the calculated phase-shifts, the control engine triggers the VNA to perform an SFCW measurement. The setup uses two horn antennas, one for transmission and one for reception, focused on the RIS surface in a quasi-monostatic configuration.

The VNA captures the complex $S_{21}$ parameters across the mmWave band, yielding a range profile for the current $(\phi_i, \theta_j)$ vector. This process is repeated for every point in the ROI. The resulting data is then aggregated into a raw 3D tensor, which undergoes further processing to generate the final voxelized volumetric image of the scene.
\subsection{Scenarios}
\begin{figure*}[!t]
\centering
\includegraphics[width=0.85\linewidth]{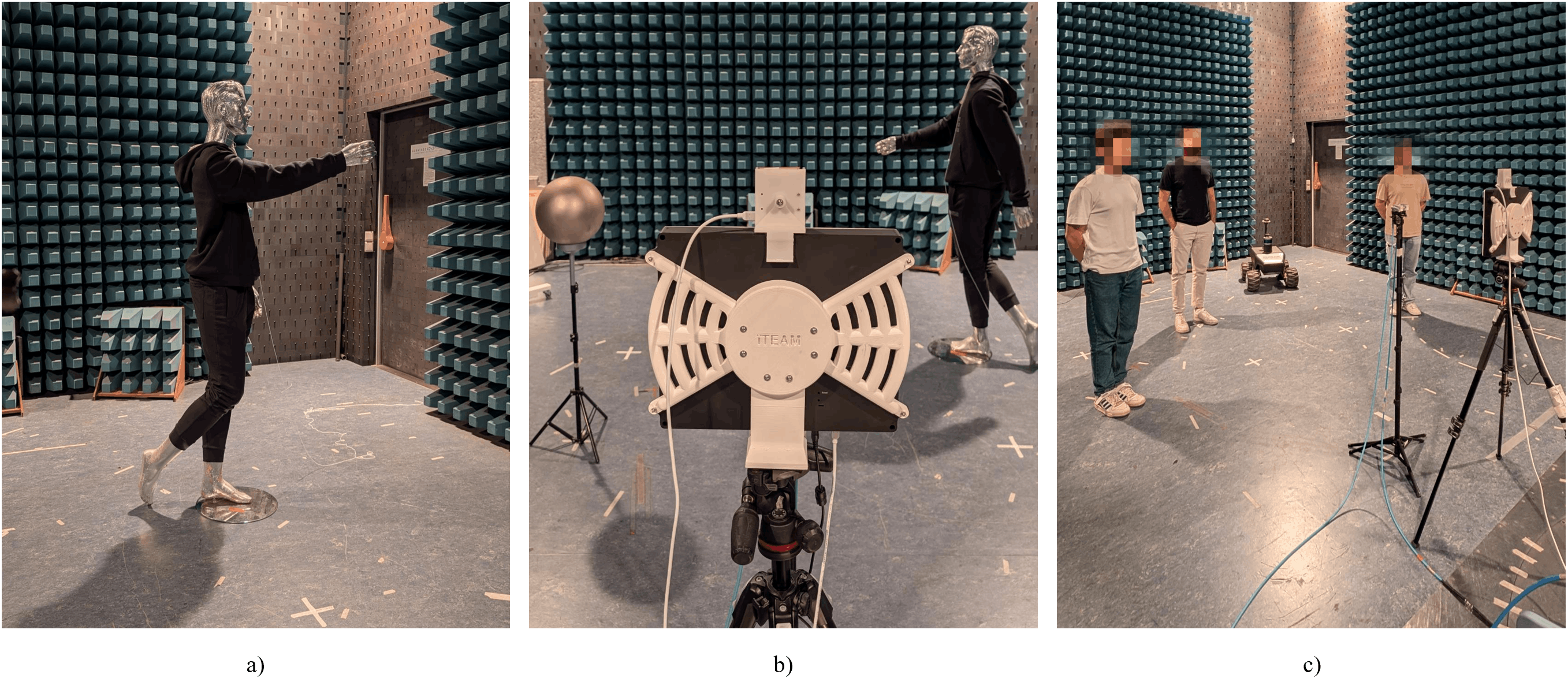}
\caption{Experimental setup for the scenarios: a) Scenario 1, single metallic mannequin; Scenario 2, metallic mannequin and sphere; Scenario 3, multi-target environment including three humans and an AGV.}
\label{fig:scenarios}
\end{figure*}

This subsection explains scenarios and the targets used in each scene. Four distinct target types were characterized: a human mannequin ($1.85$ m height, $0.50$ m width) wrapped in metallic foil to maximize reflectivity; a metallic calibration sphere ($0.15$ m radius) positioned at a mean height of $1.0$ m; three human subjects with respective heights of $1.73$ m, $1.80$ m, and $1.67$ m, all with an approximate width of $0.45$ m; and a floor-based Automated Guided Vehicle (AGV) with dimensions $0.72 \times 0.61 \times 0.42$ m ($L \times W \times H$). In each scenario, the number and type of targets were different. They are summarized in Table \ref{tab:scenarios_summary}, and images of the different scenarios are shown in Fig.~\ref{fig:scenarios}.

The system is centered at the RIS origin $(0,0,0)$, with its surface in the $YZ$-plane and the $X$-axis as its normal. The $Z$-axis is vertical, situating the ground at $z = -1.3$ m. Spherical coordinates $(R, \phi, \theta)$ are defined relative to this origin, where $\phi$ represents the azimuth in the $XY$-plane, and $\theta$ is the zenith (elevation) angle measured from the $Z$-axis.
\begin{table}[!t]
\centering
\caption{Scenarios' parameters.}
\label{tab:scenarios_summary}
\renewcommand{\arraystretch}{1.3}
\setlength{\tabcolsep}{6pt}
\begin{tabular}{|c|c|c|c|c|}
\hline
\textbf{Scenario} & \textbf{Target ID} & $\mathbf{x, y, z}~[m]$ & {$\bm{r}[m], \bm{\phi}[^\circ], \bm{\theta}[^\circ]$} & $\bm{\Delta \psi}[^\circ]$ \\ \hline
\textbf{S1} & Mannequin & 3, -1, -0.38 & 3.18, -18.4, 96.8 & 0.5\\ \hline
\multirow{2}{*}{\textbf{S2}} & Mannequin & 3, -1, -0.38 & 3.18, -18.4, 96.8 & \multirow{2}{*}{1} \\
 & Metal Sphere & 2, 1, -0.3 & 2.26, 26.6, 97.6 & \\ \hline
\multirow{4}{*}{\textbf{S3}} & Human 1 & 2, 1, -0.44 & 2.28, 26.6, 100.9 & \multirow{4}{*}{3} \\
 & Human 2 & 3, 0, -0.4 & 3.03, 0.0, 97.6 & \\
 & Human 3 & 2, -1.5, -0.47 & 2.54, -36.9, 100.5 & \\
 & AGV & 3.9, -1.20, -1.09 & 4.22, -17.1, 105.0 & \\ \hline
\end{tabular}
\end{table}

\subsubsection{Scenario 1 -- Mannequin}
This scenario is designed as the baseline for the detection-to-imaging loop. The target is a human mannequin wrapped in metallic foil to maximize Radar Cross Section (RCS). As depicted in Fig.~\ref{fig:scenarios}a, the target is isolated in the scene at an azimuth of $-18.4^\circ$. This configuration allows for the verification of the RIS beam steering accuracy and the radar's estimation precision without the interference of multipath clutter or adjacent scatterers.
\subsubsection{Scenario 2 -- Mannequin and sphere}
This scenario introduces RCS diversity and angular separation challenges. Two distinct targets are present simultaneously: the metallic mannequin and a metallic sphere. Fig.~\ref{fig:scenarios}b illustrates the RIS perspective of the mannequin and metallic ball to evaluate the system's ability to resolve two static targets separated by $45^\circ$ in azimuth.
This requires the RIS to dynamically switch focal points based on the radar detection output.
\subsubsection{Scenario 3 -- Multitarget}
The final scenario is designed as an stress test to evaluate the system's performance in the presence of multiple heterogeneous scatterers. The scene comprises three human subjects and an AGV, as shown in Fig.~\ref{fig:scenarios}c. This configuration introduces significant electromagnetic challenges, specifically due to the diversity of RCS, combining metallic and biological targets, and the increased potential for multipath propagation and mutual shadowing. The targets are distributed across a wide angular range in both azimuth and elevation, serving as a comprehensive test for the resolution and dynamic range of the imaging algorithm.
\section{System model}
\subsection{Signal Model and RIS-Aided Propagation} \label{sec:ris_phaseshift}
The imaging system utilizes an SFCW signal, defined as $K$ continuous wave (CW) pulses separated $\Delta f$, that is generated by a VNA in a quasi-monostatic configuration. Thus, the total bandwidth is $B = \Delta f(K - 1)$, and the range resolution of the system is determined by $\Delta R = \frac{c}{2B}$, where $c$ is the speed of light.

We consider a RIS with $N \times M$ passive reflecting elements. To achieve 3D imaging, the RIS is configured to focus the electromagnetic wave towards a specific point $P(x,y,z)$, in the scene. The RIS-aided propagation model is based on \cite{Tang2022PathBand}, where the target point $P$ is treated as a virtual receiver. The propagation path can be described as a dual-hop process: Transmitter (TX) $\rightarrow$ RIS $\rightarrow$ Target Point $P$ $\rightarrow$ RIS $\rightarrow$ Receiver (RX). Assuming a quasi-monostatic setup, the TX and RX paths are reciprocal.

To maximize the reflected signal from the RIS towards point $P$, the phase shifts of the RIS elements, $\psi_{n,m}$, are optimized to constructively combine the signals from all elements at that point. This is achieved by defining the optimal continuous phase shift for the $(n,m)$-th element as the negative of the total path phase:
\begin{equation}
    \psi_{n,m}^{opt} = \frac{2\pi}{\lambda} \left (r_{n,m}^{tx} + r_{n,m}^{p} \right )
\end{equation}
where $\lambda$ is the wavelength, $r_{n,m}^{tx}$ is the distance from the TX to the $(n,m)$-th RIS element, and $r_{n,m}^{p}$ is the distance from the $(n,m)$-th RIS element to the target point $P$. This formulation uses a spherical wave model, accurately accounting the spatial coordinates and distances of all points, given that the targets are located in the near-field of the RIS.

For a 1-bit RIS, the continuous phase shift $\psi_{n,m}^{opt}$ is quantized to two discrete states:
\begin{equation}
    \psi_{n,m} = \begin{cases} 0, & \text{if } \cos(\psi_{n,m}^{opt}) \geq 0 \\ \pi, & \text{if } \cos(\psi_{n,m}^{opt}) < 0 \end{cases}
\end{equation}
\subsection{Signal Processing}
The VNA measures the complex scattering parameter $S_{21}(f_k)$ for each frequency step, where $f_k = f_{start} + k \Delta f$, for $k = 0, 1, \dots, K-1$. The raw frequency-domain data is first subjected to a Hanning windowing function $W$ in the frequency domain. This is done to reduce the sidelobe levels in the final range profile, which are caused by the finite bandwidth and the rectangular pulse shape in the frequency domain. While windowing broadens the main lobe slightly, it significantly improves the dynamic range and target detectability. Finally, an Inverse Fast Fourier Transform (IFFT) is applied to the windowed $S_{21}$ data to obtain the time-domain signal, which corresponds to the range profile of the scene for the focused point $P$: $s(\phi_i, \theta_j) = \text{IFFT}\{W \cdot S_{21}(\phi_i, \theta_j)\}$. By collecting the range profile for each angle pair of the ROI, a 3D tensor is constructed as $[\phi_i, \theta_j, s(\phi_i, \theta_j)]$.
\subsection{2D Imaging and Projection}
Prior to the full volumetric reconstruction, a simplified 2D representation of the scene is generated. Given the 3D measurement tensor composed of the range profiles for each angular coordinate $(\theta_i, \phi_j)$, a 2D projection can be obtained by identifying the maximum amplitude within a specific distance window $[R_{min}, R_{max}]$. This window is dynamically adjusted based on the detections provided by the FMCW radar. The 2D intensity map $I(\theta, \phi)$ is defined as:
\begin{equation}
    I(\phi_i, \theta_j) = \max_{R \in [R_{min}, R_{max}]} |s(\phi_i, \theta_j)|,
\end{equation}
where $s(\phi, \theta)$ represents the complex range profile. This process suppresses out-of-range clutter and provides a rapid visual confirmation of the target's angular profile.
\subsection{3D Spatial Mapping and Voxelization}
The transition from angular measurements to a volumetric image relies on the spatial correspondence between the RIS beam direction and the received echoes. Due to the high-gain, narrow beams generated by the RIS, we assume that any reflection detected in the range profile at a distance $R$ originates specifically from the angular direction $(\phi_i, \theta_j)$ towards which the RIS was focused.

Thus, we define the 3D measurement signal in spherical coordinates as $S(R, \phi, \theta)$. To account for the attenuation of the signal with distance, a compensation for propagation losses is applied to each point of the raw 3D measurement signal. This results in a compensated signal $S_c(R, \phi, \theta)$, which counteracts the $1/R^4$ power loss in monostatic radar systems.

To visualize the results in a Cartesian frame, a 3D grid of voxels $\nu$ is initialized, where each element represents a discrete volume of size $l \times l \times l$ cm. The grid covers the entire hemispherical field of view of the RIS. The value assigned to each voxel at position $\mathbf{r} = (x, y, z)$ is determined by a nearest-neighbor mapping from the compensated signal $S_c$ as
\begin{equation}
\nu(\mathbf{r}) = S_c \left( \text{arg min}_{R, \phi, \theta} \| \mathbf{r} - P_{S_c}(R, \phi, \theta) \| \right),
\end{equation}
where $P_{S_c}(R, \phi, \theta)$ states for the geometrical position $(R, \phi, \theta)$ of the 3D compensated signal $S_c$.

To ensure data integrity, a distance threshold $\delta$ of $10$ cm is applied. If the distance between a voxel center and the nearest measurement point exceeds $\delta$, the voxel is assigned a null value. This prevents the interpolation of false artifacts in regions not covered by the scanning measurements.
\subsection{Volumetric Filtering and Image Generation}
The raw voxelized map often contains noise and artifacts due to side lobes and environmental scattering. To centralize the detected energy and improve the visual fidelity of the targets, a 3D Gaussian filter is applied to the voxel grid. The final filtered grid, $\nu_{f}(x, y, z)$, is defined by the convolution of the raw grid with a 3D Gaussian function $G(x, y, z, \sigma)$, followed by a non-linear thresholding operation to isolate the primary scattering centers. Let $\nu_{s}(x, y, z) = \nu(x, y, z) * G(x, y, z, \sigma)$ represent the smoothed voxel grid. The final filtered response is then expressed as:
\begin{equation}
    \nu_{f}(x, y, z) = \begin{cases} \nu_{s}(x, y, z), & \text{if } \nu_{s}(x, y, z) \geq \tau \\ 0, & \text{otherwise} \end{cases},
\end{equation}
where $\tau$ represents the intensity threshold, typically defined in the logarithmic scale to suppress the background noise floor and secondary reflections. The 3D Gaussian function $G$ used for the smoothing process is mathematically described as:
\begin{equation}
    G(x, y, z, \sigma) = \frac{1}{(2\pi)^{3/2}\sigma^3} \exp\left( -\frac{x^2 + y^2 + z^2}{2\sigma^2} \right).
\end{equation}
In this formulation, the standard deviation $\sigma$ is the critical parameter that determines the spatial extent of the smoothing. By adjusting $\sigma$, the system controls the degree of correlation between adjacent voxels, effectively merging dispersed measurements into a centralized high-power region that corresponds to the target's physical location.

This mathematical procedure, combining 2D-to-3D mapping, distance-loss compensation, and volumetric thresholded smoothing, allows for the generation of a coherent 3D volumetric imaging. The resulting reconstruction enables precise target localization and spatial profiling, providing the necessary data for tracking and scenario mapping.
\section{Results}
This section presents the imaging results obtained across three diverse experimental scenarios. The performance of the system is evaluated through two primary visualization formats: the 2D azimuth-elevation relative power map and the voxel-based 3D imaging reconstruction.
\subsection{Scenario 1}
\begin{figure}[!t]
\centering
\includegraphics[width=0.85\columnwidth]{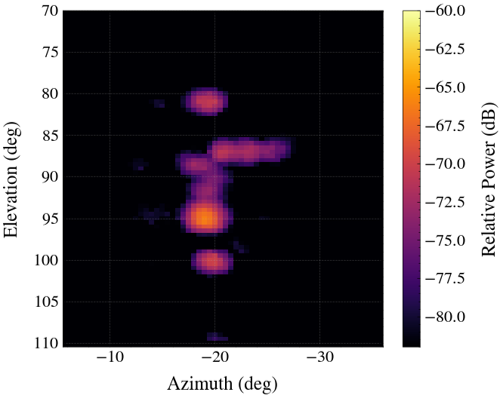}
\caption{2D Azimuth-Elevation relative power map for Scenario 1.}
\label{fig:2d_plot_mannequin}
\end{figure}

\begin{figure}[!t]
\centering
\includegraphics[width=0.9\columnwidth]{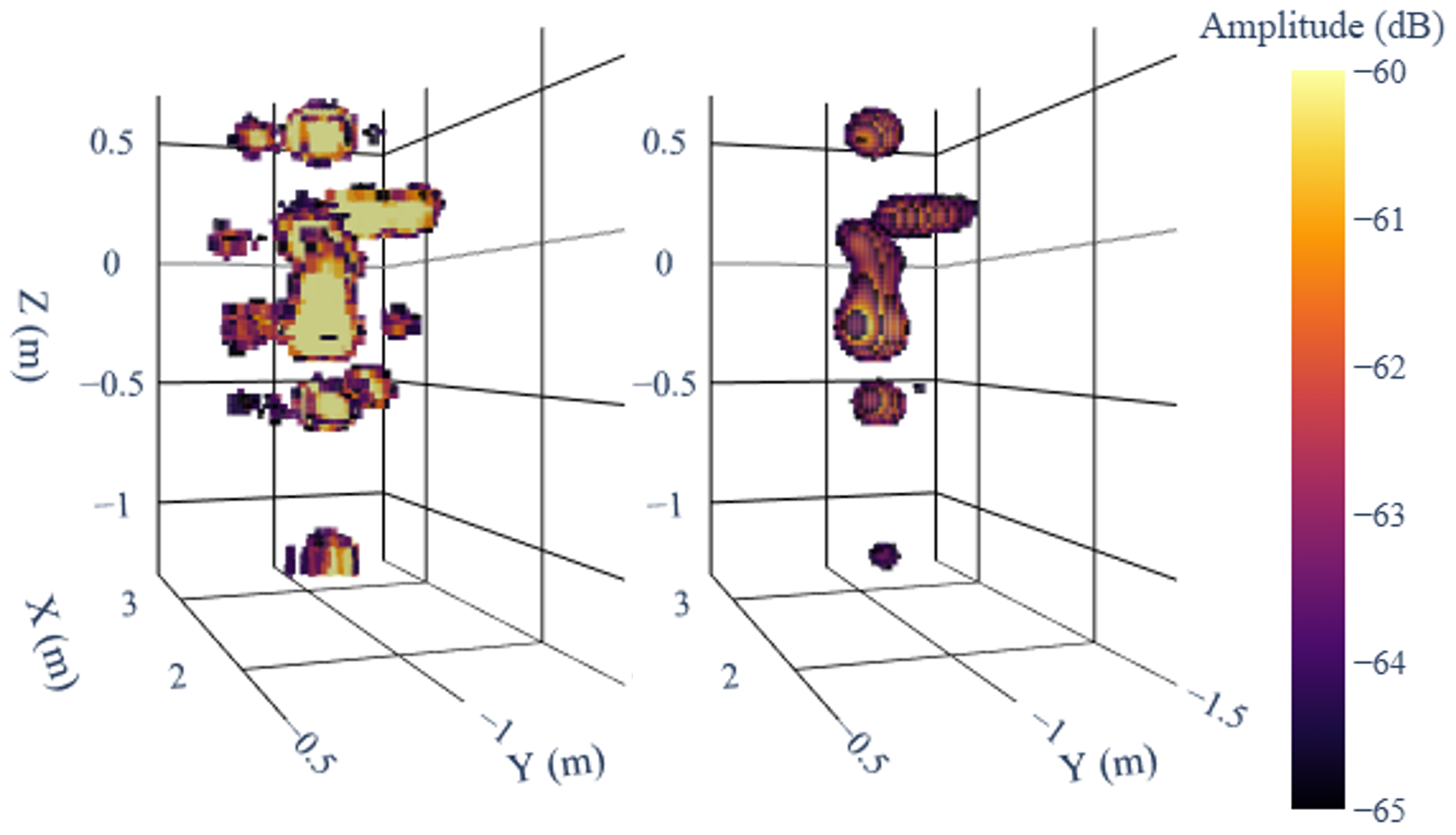}
\caption{Voxel-based 3D imaging reconstruction for Scenario 1. In the left, direct voxelization; in the right, gaussian filtering is applied.}
\label{fig:3d_plot_mannequin}
\end{figure}

In the first scenario, a metallic mannequin dressed in standard clothing is positioned in a posture with one arm extended. Fig.~\ref{fig:2d_plot_mannequin} and Fig.~\ref{fig:3d_plot_mannequin} illustrate the 2D and 3D image reconstructions, respectively.

The 2D projection accurately identifies the mannequin's angular position, showing a high-intensity silhouette corresponding to the torso and the extended arm. The 3D imaging shows on the left the result without gaussian filtering, where noise and shapes are disrupted; while in the right, when gaussian filtering is applied, it provides a faithful representation of the mannequin's posture, location, and volumetric shape. The torso, head, and extended arm are clearly discernible. However, certain regions, such as the neck and portions of the legs, are less visible. This effect is attributed to the surface normals in these areas not being oriented toward the RIS, leading to electromagnetic wave scattering away from the receiver. Despite these minor discontinuities, the overall 3D envelope and posture are successfully captured.
\subsection{Scenario 2}
\begin{figure}[!t]
\centering
\includegraphics[width=0.85\columnwidth]{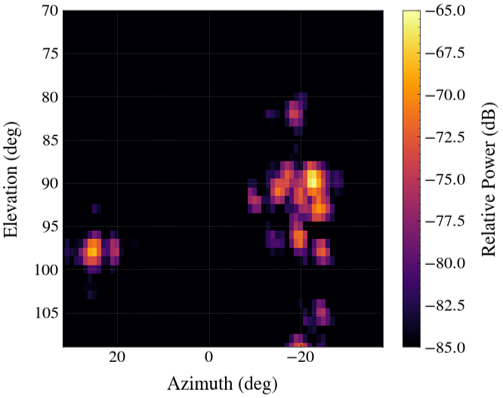}
\caption{2D Azimuth-Elevation relative power map for Scenario 2.}
\label{fig:2d_plot_mannequin_and_sphere}
\end{figure}

\begin{figure}[!t]
\centering
\includegraphics[width=0.9\columnwidth]{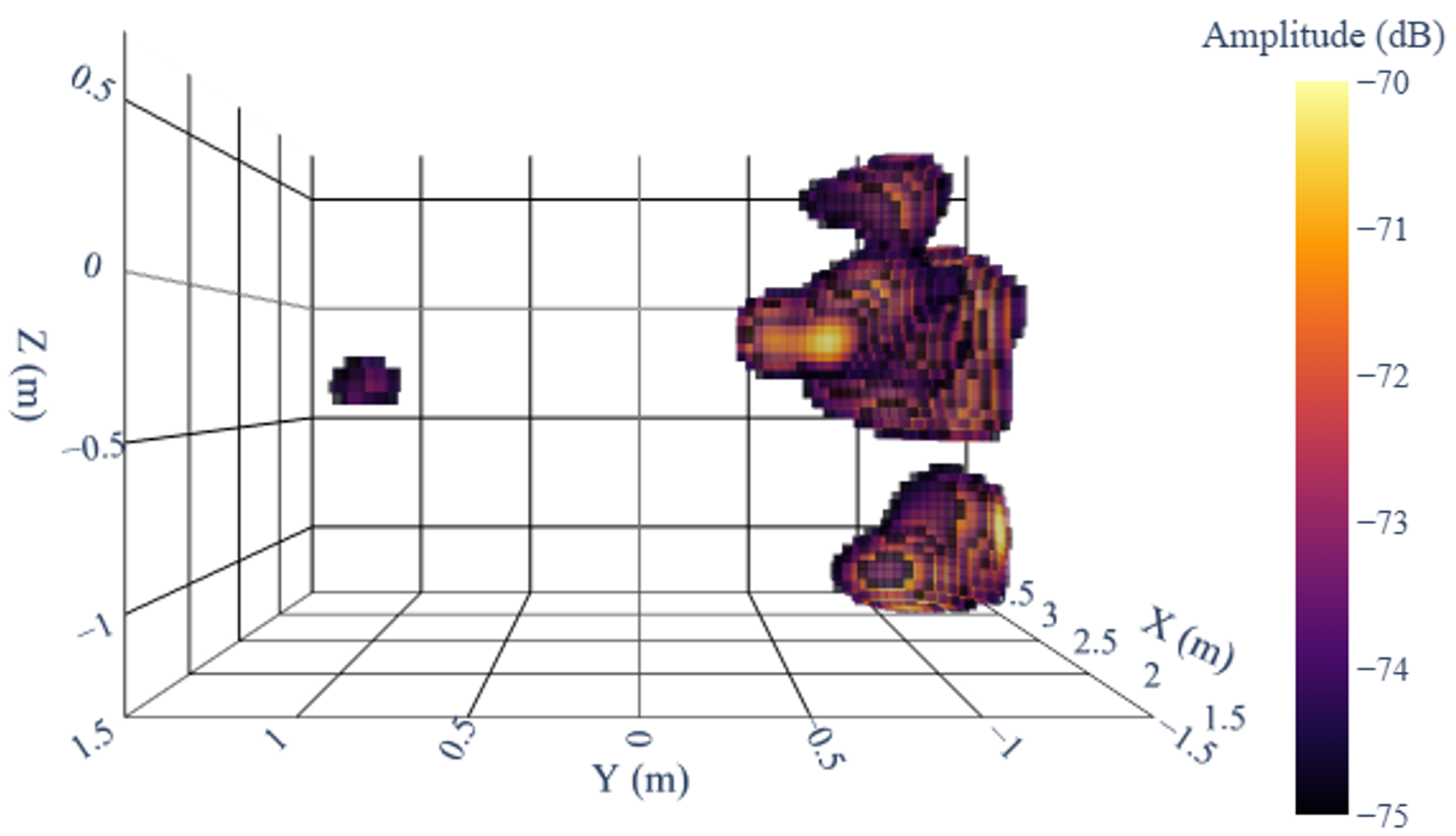}
\caption{Voxel-based 3D imaging reconstruction for Scenario 2.}
\label{fig:3d_plot_mannequin_and_sphere}
\end{figure}

Scenario 2 involves a more complex configuration consisting of a metallic sphere and the metallic mannequin in a walking posture. Fig.~\ref{fig:2d_plot_mannequin_and_sphere} and Fig.~\ref{fig:3d_plot_mannequin_and_sphere} are the 2D and 3D imaging reconstructions, respectively for the Scenario 2.

In this scenario, both targets are localized at their correct elevation and azimuth coordinates. The intensity map for the mannequin begins to reveal the walking posture, while the sphere appears as a concentrated high-power spot. The volumetric reconstruction shows a distinct, nearly spherical shape at the sphere's location. For the mannequin, the 3D map captures the head, torso, and extended arm. While the proximity between the limbs and the torso makes it challenging to separate individual extremities, the resulting voxel grid effectively represents the enveloping volume of the walking posture. The measurements demonstrate that the system can distinguish between different geometric types (spherical vs. humanoid) within the same measurement cycle.
\subsection{Scenario 3}
The final scenario represents a challenging stress test for the system, featuring an AGV and three human subjects at varying distances and angles. For this scenario, only the Voxel-based 3D imaging reconstruction is analyzed, as the 2D maximum-intensity projection fails to effectively separate targets and suppress noise due to the significant diversity in the distance, angle and reflectivity of the various targets.

\begin{figure}[!t]
\centering
\includegraphics[width=0.9\columnwidth]{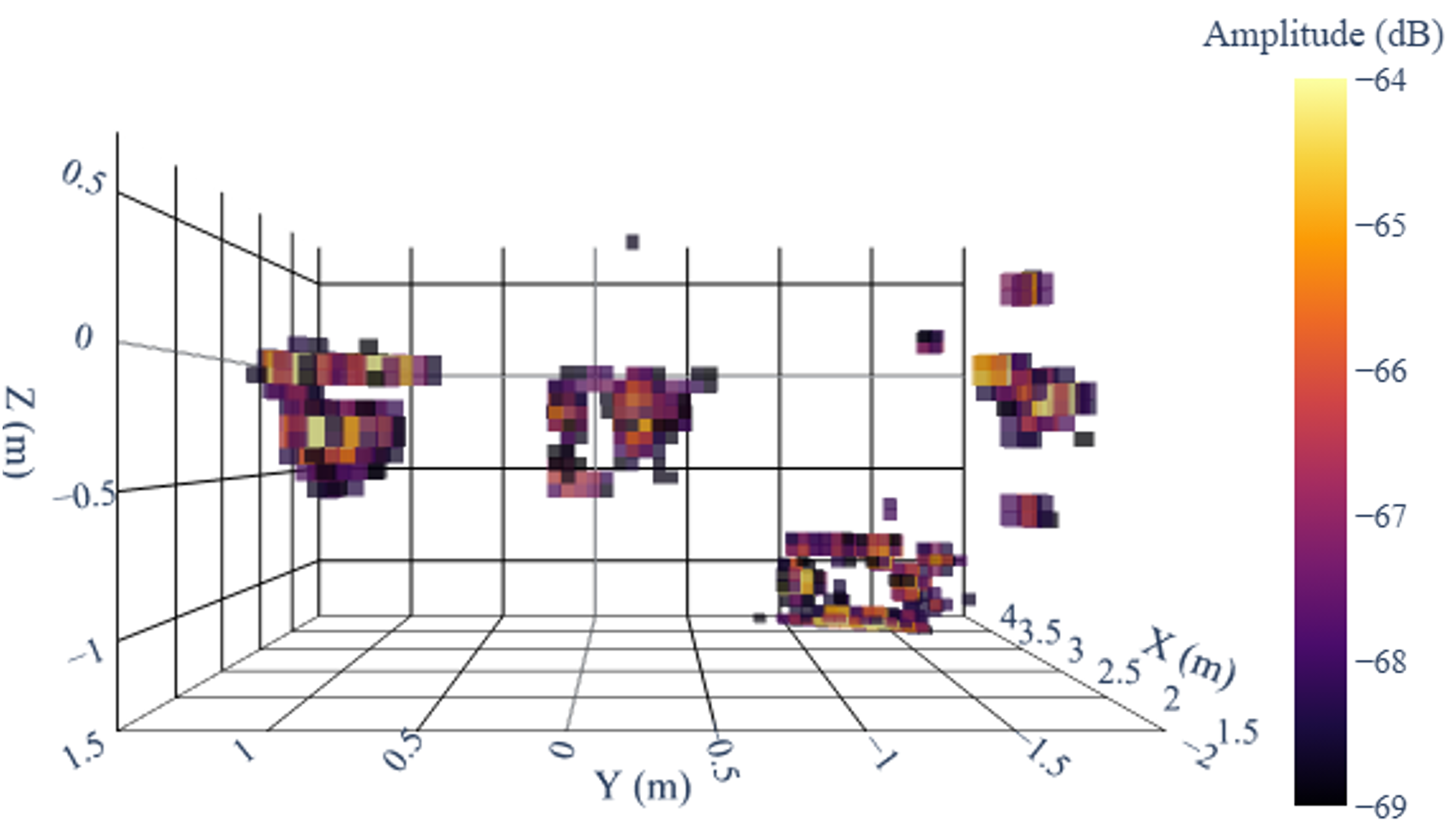}
\caption{Voxel-based 3D imaging reconstruction for Scenario 3.}
\label{fig:3d_plot_mult_targets}
\end{figure}

In this scenario, there are some system challenges, such as humans who present a significantly lower RCS compared to metallic targets, making detection more difficult. Furthermore, to test the system's operational limits, the angular step $\Delta \psi$ between beams was increased to $3^\circ$ to limit the angular resolution. The 3D voxelized imaging reconstruction of the scenario is shown in Fig.~\ref{fig:3d_plot_mult_targets}.

The 3D image shows the AGV reconstruction being located in the lower-right area and furthest from the RIS. Its geometric box-like shape is visible, and even the cylindrical components on its roof are discernible in the grid. Also, the three human subjects are correctly localized in the 3D space. Due to their lower RCS, the reconstruction focuses on the high-scattering regions: the torso and shoulders are visible for all three, while the heads are represented by smaller clusters of voxels, particularly for the subjects in the center and right of the scene.

Despite the coarse angular resolution (due to the $3^\circ$ step), the processed 3D voxel image enables target localization and provides sufficient morphological information to distinguish the target type (e.g., the rigid structure of an AGV vs. the complex silhouette of a human). This validates that the RIS-based imaging system is robust even in multi-target, low-RCS and low-angle-resolution scenarios.

Finally, the parameters of the 3D imaging such as the voxel size ($l$) and the Gaussian fitting ($\tau,\sigma$) performed in the different scenarios, are shown in Table~\ref{tab:3dparams}.
\begin{table}[!t]
\centering
\caption{3D imaging parameters.}
\label{tab:3dparams}
\renewcommand{\arraystretch}{1.2}
\begin{tabular}{|c|c|c|c|}
\hline
\textbf{Parameter} & \textbf{Scenario 1} & \textbf{Scenario 2} & \textbf{Scenario 3} \\ \hline
\textbf{$l$} & 2~cm & 4~cm & 6~cm \\ \hline
\textbf{$\tau$} & -65~dB & -79~dB & -85~dB  \\ \hline
\textbf{$\sigma$} & 2 & 2 & 0.8 \\ \hline
\end{tabular}
\end{table}

% The standard deviation $\sigma$ is maintained for scenarios featuring metallic targets, specifically when the mannequin is present. However, the heterogeneity of the third scenario, which involves humans and an AGV, necessitates a distinct $\sigma$ value to accommodate the diverse scattering properties of the subjects.
\section{Discussion}
The described experimental work relies on a novel RIS-aided 2D and 3D imaging system at mmWave frequencies combined with quasi-monostatic SFCW radar. 
%This work demonstrates a novel experimental RIS-aided 2D and 3D imaging system at mmWave frequencies using a quasi-monostatic SFCW radar. 
The primary innovation lies in the autonomous, closed-loop operation of the experimental setup. By integrating an FMCW radar to serve as the "eyes" of the RIS, the system dynamically updates its sensing cycle, identifying targets and defining ROIs in real-time as the environment changes. This synergy allows for a significant reduction in measurement time by focusing high-resolution scanning only on areas of interest, proving that the RIS can be transformed from a communication-centric device into a powerful, adaptive sensing tool.

Despite the effective reconstructions, some hardware-related constraints were identified. Firstly, the performance is sensitive to the operational bandwidth of the RIS hardware. Optimal $180^\circ$ phase shifts between states are typically achieved within a limited frequency range, affecting the signal-to-noise ratio (SNR) in wideband measurements. Secondly, the use of a 1-bit phase resolution per polarization introduces quantization errors. This leads to slight pointing inaccuracies, where the generated beams may not align perfectly with the intended 3D coordinates. Thirdly, the current setup utilizes only one of the two available linear polarizations of the RIS. Consequently, simultaneous V and H linear polarizations or cross-polarized scattering information, which could enhance target classification, is currently not exploited.

To overcome these limitations, some enhancements are proposed. On the one hand, increasing the number of bits for phase control would provide finer beam steering and significantly reduce quantization artifacts. Furthermore, scaling the RIS to a larger number of elements would enable narrower beams, further improving spatial resolution. On the other hand, future iterations could implement a dual-polarization multiplexing scheme. By utilizing both linear polarizations simultaneously, the system could capture more complex polarimetric signatures of the targets, improving the reconstruction of low-RCS objects like humans. In addition, deploying multiple, delocalized RIS units would allow for a multi-perspective volumetric representation. Finally, employing a monostatic FMCW radar as the RIS feeder for the range profile obtention would enable real-time 3D imaging. In particular, commercial FMCW radars can generate thousands of chirps per second. Given that the current RIS hardware supports an update frequency of at least 1 kHz, with a maximum of 75 kHz, the system could realistically process thousands of beams per second. This speed would enable not only static real-time 3D imaging but also high-speed 3D tracking of multiple moving targets within the scene.
\section{Conclusion}
This paper has successfully presented and experimentally validated an autonomous mmWave imaging framework that exploits the synergy between a commercial FMCW radar and a 1-bit RIS. By implementing a closed-loop control architecture, the system uses the radar as a spatial guide to dynamically identify ROIs, focusing the high-resolution sensing capabilities of the RIS only where required. This approach enables targeted 3D measurements and spatial mapping, effectively overcoming the limitations of exhaustive codebook scanning and significantly improving the scalability and practical feasibility of RIS-based imaging systems. Furthermore, experimental results demonstrate that the proposed 3D voxelization framework, enhanced by distance-loss compensation and volumetric Gaussian filtering, provides effective reconstructions of heterogeneous targets, including complex metallic structures and human subjects with low RCS.

The findings indicate that while hardware constraints such as 1-bit quantization and limited RIS bandwidth introduce slight pointing inaccuracies, the system remains robust across various environmental complexities. Future work may focus on scaling the architecture through multi-bit RIS designs for finer beam steering, dual-polarization multiplexing to enhance target classification, and transitioning to a fully monostatic FMCW-based feeder to enable high-speed, real-time 3D-image tracking. In this context, the demonstrated closed-loop operation provides a viable pathway toward time-efficient, adaptive sensing with intelligent surfaces, aligning with the dynamic requirements of future 6G ISAC ecosystems.
\section*{Acknowledgment}
This work has been funded by the COREMAT-6G projects (TSI-063000-2021-118, TSI-063000-2021-119, TSI-063000-2021-120) of the UNICO-5G R\&D 2021 Programme of the PRTR, funded by the European Union NextGenerationEU (PRTR- C15.I06) with the support of the Ministry of Economic Affairs and Digital Transformation - Government of Spain.



\bibliographystyle{IEEEtran}

\bibliography{config,references}  

\end{document}